\documentclass[mypaper,7pt,twoside]{CoAst}
\usepackage{epsf,graphicx,fancyhdr}
\renewcommand{\chaptermark}[1]{\markboth{#1}{}}

\newcommand{\kopf}{\small\itshape Comm. in Asteroseismology \\ Contribution to the Proceedings of the Wroclaw HELAS Workshop, 2008}

\newcommand{\Authors}[1]{\begin{center}\normalsize\bf\sf #1 \end{center}}

\renewcommand{\author}[1]{\begin{center}\normalsize\bf\sf #1 \end{center}}
\newcommand{\Address}[1]{\begin{center}\small\sf #1 \end{center}}

\newcommand{\Session}[1]{{\vspace{3mm}\small \noindent  \hspace*{3mm} Session: } #1 \normalsize}

\newcommand{\Objects}[1]{{\vspace{0mm}\small \noindent  \hspace*{3mm} Individual Objects: } \small #1 \normalsize}

	\newcommand{\oneA}{\small DATA - ground-based photometry \newline}

\renewenvironment{abstract}{\section*{Abstract}\normalsize\sf}{}
\newcommand{\References}[1]{\begin{flushleft}{\large References\\}\vspace*{2mm}\small #1 \end{flushleft}}

\newcommand{\chapterCoAst}[2]{\chapter[\sf\normalsize #1\\ \footnotesize \hspace*{5mm}by #2 \sf\normalsize][]{#1\\}\rhead[\fancyplain{}{\sf\footnotesize \center{#1}}]{\fancyplain{}{\sffamily\thepage}}\lhead[\fancyplain{\kopf}{\sffamily\thepage}]{\fancyplain{\kopf}{\sf\footnotesize \center{#2}}}}

\newcommand{\figureCoAst}[5]{\begin{figure}[#4]
\centering
\includegraphics*[#5]{#1}
\caption{#2}
\label{#3}
\end{figure}}

\renewcommand{\chaptername}{}

\def\rfr{\smallskip\par\noindent
        \hangindent=7truemm
        \hangafter=1}

\begin{document}
\sf

\chapterCoAst{Multi-wavelength photometric variation of PG\,1605$+$072}
	     {S.~Schuh, S.~Dreizler, U.~Heber, et.\ al}
\Authors{S.~Schuh$^{1}$, S.~Dreizler$^{1}$, U.~Heber$^{2}$, 
  C.S.~Jeffery$^{3}$, S.J.~O'Toole$^{4,2}$, O.~Cordes$^{5}$, 
  T.~Stahn$^{6,1}$, R.~Lutz$^{1,6}$, A.~Tillich$^{2}$, 
  and the WET and MSST collaborations} 
\Address{
  $^{1}$Institut f\"ur Astrophysik, Georg-August-Universit\"at G\"ottingen,
  Friedrich-Hund-Platz~1, 37077 G\"ottingen, Germany\\
  $^{2}$Dr.~Remeis-Sternwarte Bamberg, Universit\"at Erlangen-N\"urnberg,\\
  Sternwartstra\ss e~7, Germany\\
  $^{3}$Armagh Observatory, College Hill, Armagh BT61 9DG,\\
  Northern Ireland, United Kingdom\\
  $^{4}$Anglo-Australian Observatory, P.O. Box 296, Epping, NSW 1710,
  Australia\\
  $^{5}$Argelander-Institut f\"ur Astronomie, Universit\"at Bonn,\\
  Auf dem H\"ugel 71, 53121 Bonn, Germany\\
  $^{6}$Max-Planck-Institut f\"ur Sonnensystemforschung, Max-Planck-Stra\ss e 2,\\
  37191 Katlenburg-Lindau, Germany
}
\noindent
\begin{abstract}
In a large coordinated attempt to further our
understanding of the $p$-mode pulsating sdB star
\mbox{PG\,1605$+$072}, the Multi-Site Spectroscopic Telescope (MSST)
collaboration has obtained simultaneous time-resolved
spectroscopic and photometric observations. The
photometry was extended by additional WET data which
increased the time base. This contribution outlines the analysis of
the MSST photometric light curve, including the
four-colour BUSCA data from which chromatic amplitudes
have been derived, as well as supplementary FUV spectra and light
curves from two different epochs. These results
have the potential to complement the interpretation of
the published spectroscopic information.
\end{abstract}
\par
\Session{ \oneA }
\Objects{PG\,1605$+$072} 
\section*{Multi-Site Spectroscopic Telescope for PG\,1605+072}
\subsection*{Introduction and MSST overview} %
\mbox{PG\,1605+072} is a pulsating subdwarf B star evolving off the
EHB. Observationally, it may be considered a sibling among
sdBs of the even brighter star \mbox{Balloon~090100001}; both stars
are 
multiperiodic and pulsate each with the largest amplitude among the
sdBs of $\approx$6\% in the strongest mode. Just as more recently for
\mbox{Balloon~090100001}, the rich frequency spectrum has triggered
extended photometric monitoring campaigns in the optical as well as 
the gathering of time-resolved colour and spectral information.
The resulting literature that has been published, from the initial
discoveries and campaigns all the way to the work in progress on analysing the
repeated coordinated observational efforts, is too numerous to be cited
comprehensively in this context. 
\par
The Multi-Site Spectroscopic Telescope (MSST) project in particular
combined the following observational ingredients in order to
simultaneously sample \mbox{PG\,1605$+$072}'s intensity and radial
velocity variations: white light and 
multicolour light curves;
low-resolution time-resolved spectroscopy
(O'Toole et al.\ 2005, Tillich et al.\ 2007);
and high-resolution time-resolved spectroscopy.
\par
This report primarily makes mention of the photometric analysis, 
and furthermore discusses \mbox{PG\,1605$+$072} as observed with FUSE:
light curves, radial velocities, and especially chromatic amplitudes
as presented by Stahn 2005 and Lutz 2007.
\par 
In the spirit of this workshop, the focus is on the variety of data
sets available and how to treat and potentially combine this data.
For published results (in numbers) the relevant work is referenced;
here we note that the immediate aim of MSST is mode identification
($l$) to complement future asteroseismic modelling. The changing power
in the pulsation spectrum of \mbox{PG\,1605$+$072} may make this
difficult, but may also hold clues to the details of the driving
mechanism. The long-term motivation remains the clarification of the
evolutionary status and origin of \mbox{PG\,1605$+$072} as one
representative of the subdwarf~B stars.
\subsection*{MSST and WET optical light curve, Fourier transforms and frequency fitting}
The optical light curve consists of a total of 96 individual data
sets, combining white light data from the MSST photometry and the 
WET Xcov22 campaign. The frequency solution for the white light curve has
been obtained in an iterative manner. First of all, overlapping data
sets were cross-correlated to check the timing and quality. On a
trusted subset of the data obtained by bootstrapping from the
overlapping data sets, first
a four-, later an eleven-frequency model was constructed. The initial
model was cross-correlated with all observations to uncover and correct
for remaining timing errors, improved using the provisionally
corrected full data set, and the procedure repeated with the second,
more complex model. Finally, a 55-frequency model was fitted to the
corrected data, using a non-linear least squares sine fit as in the
steps before.
\par
The final light curve documents this procedure by
providing, for each data point, the time as raw truncated Julian Date,
followed by the value of the barycentric correction, the time in BJD,
the value of the corrections derived from the cross correlation, and
finally the corrected time in BJD. This is followed by the modulation
intensity with a mean value of zero, and an observation ID unique to
each of the 96 data sets.
\section*{Chromatic amplitudes}
\subsection*{MSST: BUSCA multicolour light curve}
\figureCoAst{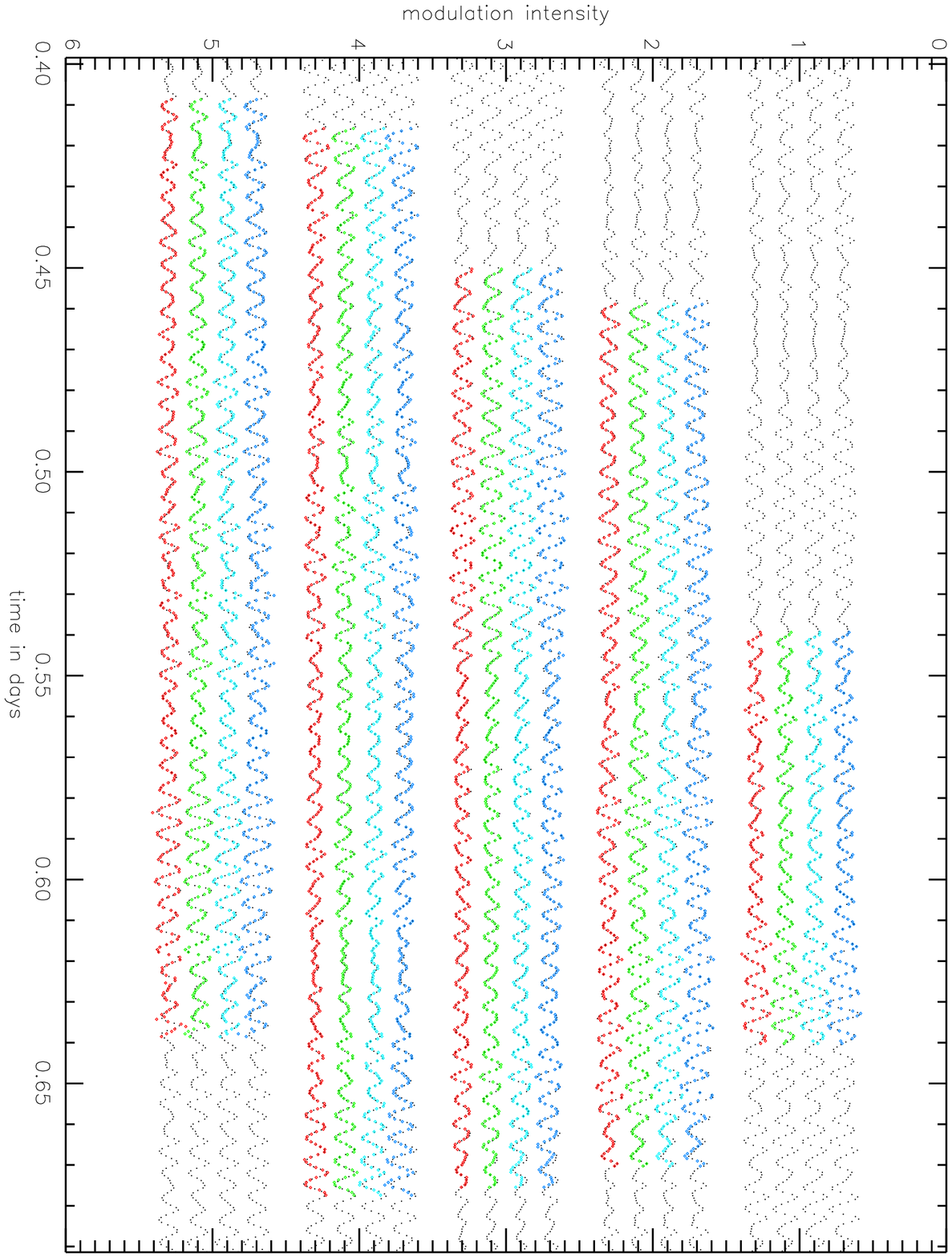}{
  Five nights of multicolour photometry obtained in 2002 with the
  BUSCA instrument attached to the Calar Alto 2.2m telescope. The data points
  in the top four curves correspond (in descending order) to the unfiltered
  \texttt{uv}, \texttt{b}, \texttt{r}, \texttt{nir} light curves of
  May 14, followed by the data sets corresponding to May 18, 19, 20, and
  21. A continuous model light curve containing 30 frequencies is overplotted.
  }{busca}{!t}{clip,angle=90,width=\textwidth}
\figureCoAst{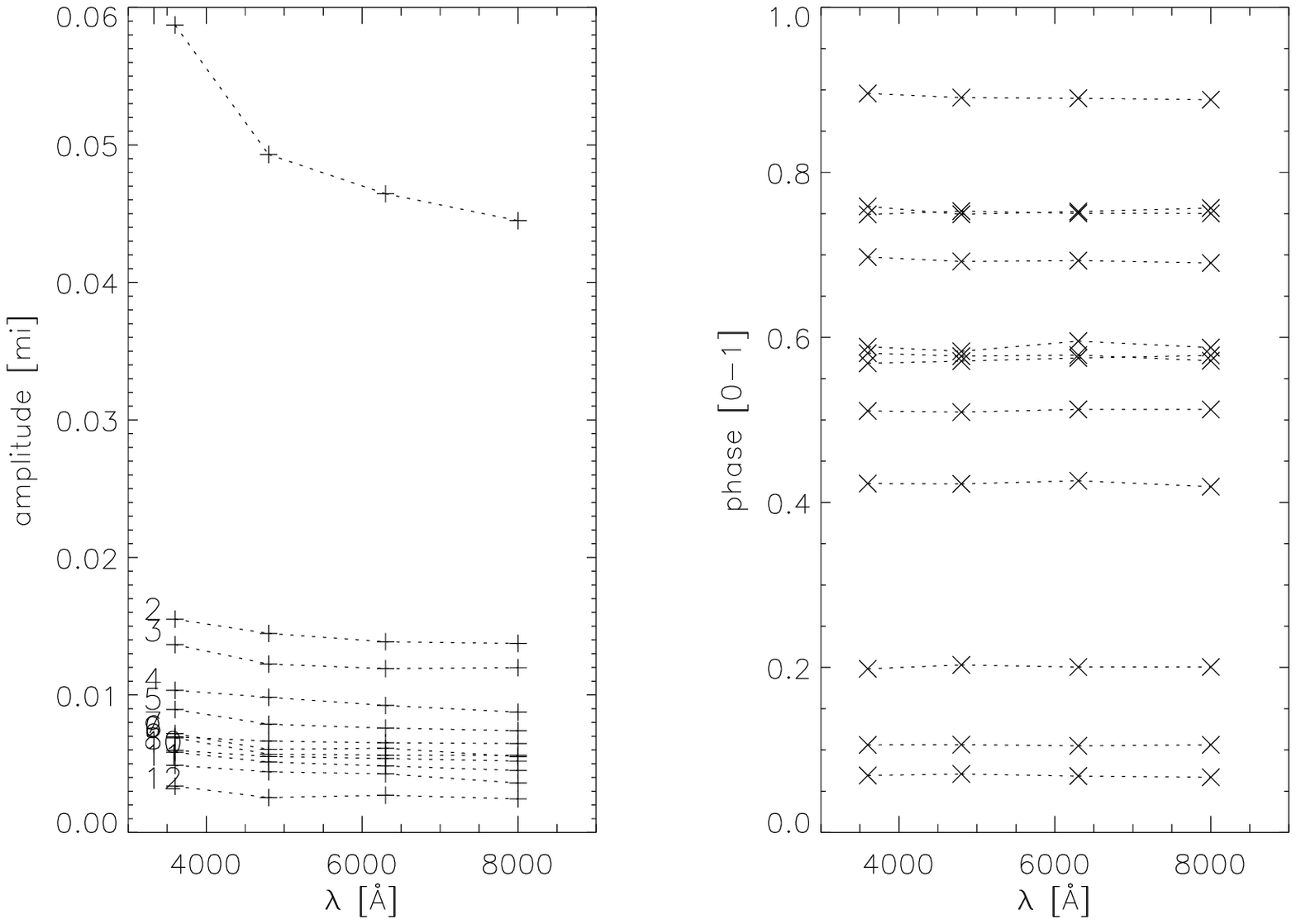}{
  \emph{Left panel:} Chromatic amplitudes derived from the 2002 BUSCA data shown
  in Fig.\,\ref{busca} for the twelve strongest frequencies in the
  model light curves. The error bars correspond roughly to the symbol
  size; beyond the 12${\textrm{th}}$ strongest frequency, error bars
  are larger than the measurable slope.
  \emph{Right panel:} A quality check of the fit is obtained by
  inspecting the stability of the phase (free parameter; arbitrary
  zero point). 
}{chromatic}{!t}{clip,angle=0,width=\textwidth}
Tremblay et al.\ 2006 have compiled and analysed the optical
multicolour photometry available for pulsating sdB stars at the
time. This included the measurements by Falter et al.\ 2003 obtained
in 2001 for \mbox{PG\,1605$+$072}. Figure~\ref{busca} shows the five
nights of multicolour photometry obtained in 2002 with
the BUSCA instrument at the Calar Alto 2.2m telescope. To construct
the overplotted model, the BUSCA colours were first collapsed into one
''white'' light curve, which was fitted with a reduced set of
frequencies from the simultaneous, more extended white light MSST
photometry. The reduction of the number of frequencies was done by
merging close frequencies, not resolved in the BUSCA data subset,
assuming they belong to the same $l$. This may not be correct; if so,
the chromatic amplitudes derived for these 
merged frequencies will be meaningless. Chromatic amplitudes were
obtained by fixing the 30 frequencies fitted to the BUSCA ''white''
light curve and re-determining amplitudes and phases on the individual
\texttt{uv}, \texttt{b}, \texttt{r}, and \texttt{nir} light
curves. The result for the twelve strongest frequencies is shown in
Fig.\,\ref{chromatic}. In a plot where the chromatic amplitudes are
normalised to the \texttt{uv} amplitude (not shown), the curves
roughly fall into three groups with different slopes, indicative of
the expected grouping according to common $l$ values.
\subsection*{FUSE far-UV light curve}
As shown by Fontaine \& Chayer 2006, far-UV light curves for
\mbox{PG\,1605$+$072} may be obtained by collapsing FUSE spectra
extracted in bins from time-tagged data producing a time series.  If
the collapsing process is applied to a subset of wavelength bins,
chromatic amplitudes for the far-UV spectral range can be determined
from the resulting individual intensity time series. This has been
done by Stahn 2005 for the 07/2001 data (see also Lutz 2007 for the
04/2004 data). Again, a smaller subset of merged frequencies was used
when deriving the chromatic amplitudes, with close frequencies not
resolved in the FUSE Fourier spectrum merged together. The observed wavelength dependency of the pulsation amplitudes was compared to model
predictions, but did not allow a reliable mode identification. 
\section*{Radial velocities and spectroscopic parameters}
\subsection*{FUSE radial velocities}
The time series of uncollapsed FUSE spectra can be subjected to a
cross-correlation analysis which yields pulsational radial
velocities. The 07/2001 radial velocity curve has been analysed by
Stahn 2005 in a similar way to the analysis done on the FUV intensity.
In a direct comparison of the full curves, Stahn 2005 finds that the
light to radial velocity phase shift amounts to $\pi$/3, a result
which differs from that published by Kuassivi et al.\ 2005 who find
$\pi$/2. When the comparison is done for individual frequencies, Stahn
2005 notices that the $\pi$/3 phase shift basically reflects the value
corresponding to the strongest frequency, and derives differing values
for two further frequencies. These results are indicative of a
non-adiabatic pulsational behaviour of \mbox{PG\,1605$+$072}.
\subsection*{Optical radial velocities and spectroscopic parameters}
O'Toole et al.\ 2005 derived the radial velocity variation from MSST
low-resolution spectra. These results were used by Tillich et al.\
2007 to produce RV-corrected, phase-resolved summed spectra for the
strongest pulsation frequencies from the same optical
spectroscopy. From these the variation in effective temperature and
surface gravity corresponding to the individual pulsations could be
determined. 
Phase relations were obtained for the photospheric parameters radial
velocity versus $\log{g}$ and radial velocity versus temperature
variation. As expected from geometrical considerations, the radial
velocity versus $\log{g}$ variation results in a shift of $\pi/2$. The
shift in radial velocity versus temperature variation of $\pi/3$ is
fully consistent with the FUSE results if the intensity variation is
assumed to be primarily due to changes in the effective temperature. 
Again, this points to the presence of non-adiabatic pulsational behaviour. 
\subsection*{Challenges}
To fully exploit the available observations briefly presented above,
the following exercises remain to be carried out on the data.
First, an in-depth analysis of the white light curve should be able to
discriminate from the behaviour of phases if either genuine amplitude
variations or instead unresolved modes are seen. This will also be of
relevance for justifying the merging of frequencies when determining 
chromatic amplitudes; here the 2002 BUSCA optical chromatic amplitude
results need still to be compared to Falter et al.\ 2003 (2001 data).
The same argument holds for the analysis of the short FUSE data sets
that do not fully resolve the pulsation spectrum; some improvement can
be expected when adding in the second (2004) FUSE data set. The
challenge will then be to bring together the non-simultaneous
optical and FUV chromatic amplitudes; the latter can in principle
provide a very significant lever. 
\par
The next interesting step will be to bring together the optical light
curves and the spectroscopy: will the same phase lags as found in the 
FUV, and similarly indicated by the variation in the spectroscopic
parameters, be directly evident there? It also remains to be seen if
the overall line profile variations will be matched by modelling
attempts. Finally, it will be interesting to find out if the mode
identification with these methods continues to remain a 
fundamental challenge despite the encouraging intermediate results.
\par
\References{
\rfr Falter, S., Heber, U., Dreizler, S., et al.\ 2003, A\&A, 401, 289
\rfr Fontaine, G., \& Chayer, P. 2006, ASPC, 348, 181
\rfr Lutz, R. 2007, diploma thesis, University of G\"ottingen
\rfr O'Toole, S. J., Heber, U., Jeffery, C. S., et al.\ 2005, A\&A, 440, 667
\rfr Kuassivi, Bonanno, A., \& Ferlet, R. 2005, A\&A, 442, 1015
\rfr Stahn, T. 2005, diploma thesis, University of G\"ottingen
\rfr Tillich, A., Heber, U., O'Toole, S. J., et al.\ 2007, A\&A, 473, 219
\rfr Tremblay, P.-E., Fontaine, G., Brassard, P., et al.\ 2006, ApJS,
  165, 551
}
\end{document}